\documentclass[twocolumn,superscriptaddress,prl]{revtex4}
\usepackage{amsmath}

\usepackage{graphicx}
\usepackage{dcolumn}
\usepackage{bm}
\usepackage{xcolor} 
\usepackage[normalem]{ulem} 
\newcommand{\ML}{}
\newcommand{\stkout}[1]{}

\newcommand{\kT}{k_{\rm B}T}
\newcommand{\md}{\mathrm{d}}
\newcommand{\st}{\boldsymbol{\sigma}} 
\newcommand{\logit}{f} 

\begin{document}
	
\preprint{APS/123-QED}
	
\title{Connections between efficient control and spontaneous transitions in an Ising model}
	
\author{Miranda D.\ Louwerse}
\affiliation{%
	Department of Chemistry, Simon Fraser University, Burnaby, British Columbia, Canada V5A1S6
}%
\author{David A.\ Sivak}
\email{dsivak@sfu.ca}
\affiliation{%
	Department of Physics, Simon Fraser University, Burnaby, British Columbia, Canada V5A1S6
}%

\date{\today}
	
\begin{abstract}
	A system can be driven between metastable configurations by a time-dependent driving protocol, which uses external control parameters to change the potential energy of the system. Here we investigate the correspondence between driving protocols that are designed to minimize work and the spontaneous transition paths of the system in the absence of driving. We study the spin-inversion reaction in a 2D Ising model, quantifying the timing of each spin flip and heat flow to the system during both a minimum-work protocol and a spontaneous transition. The general order of spin flips during the transition mechanism is preserved between the processes, despite the coarseness of control parameters that are unable to reproduce more detailed features of the spontaneous mechanism. Additionally, external control parameters provide energy to each system component to compensate changes in internal energy, showing how control parameters are tuned during a minimum-work protocol to counteract underlying energetic features. This study supports a correspondence between minimum-work protocols and spontaneous transition mechanisms.
\end{abstract} 
	
	
\maketitle

\section{Introduction}

Quantifying the dynamics and energetics of a system as it undergoes a spontaneous transition between metastable states is of interest to the natural sciences due to the ubiquity of such activated processes throughout chemistry and biology~\cite{Karplus2002,Chodera2014,E2010}. The system typically overcomes a free-energetic barrier separating metastable states in its high-dimensional configuration space, requiring collective motion of many degrees of freedom and heat flow from the environment to increase the system's internal energy (in reactions with an energy barrier). Characterizing the thermodynamics and kinetics of the collective variables involved in the motion is therefore of interest~\cite{Peters2016,Peters2017q,Bolhuis2000,Wang2019}. A system can also be driven through its configuration space through time-dependent variation of external control parameters that provide an energetic bias to (sets of) collective variables. Excess work (work above the equilibrium free-energy change) is done on the system during a protocol depending on how the system is driven, making it a target for optimization~\cite{Schmiedl2007,Sivak2012}. Here, we investigate the correspondence between driving protocols that minimize work in the long-duration limit and the spontaneous transition mechanism through configuration space\stkout{.}{\ML, hypothesizing that minimum-work protocols effectively make use of spontaneous fluctuations by providing work to each degree of freedom in accordance with its required heat intake during a spontaneous transition.}

Driving protocols can be implemented in experiment and simulation~\cite{Liphardt2002,Engel2014,Dudko2008,Gupta2011,Morfill2008,Moradi2014}; coupled with theoretical advances~\cite{Jarzynski1997,Crooks2000,Hummer2001}, driving protocols are a widely applicable tool for extracting equilibrium thermodynamic information about a variety of microscopic systems. The ability to estimate equilibrium properties is perhaps surprising since the system is out of equilibrium throughout the driving protocol, and therefore kinetic aspects of the system's response to control-parameter perturbations are highly relevant~\cite{Hummer2001,Mandal2016,Schmiedl2007}. 

The excess work on the system during a driving protocol performed in long duration can be approximated by linear-response theory~\cite{Sivak2012}, yielding a geometry in control-parameter space with a generalized friction metric that quantifies the system's resistance to changes in control parameters. This approximation also yields an intuitive description of minimum-work protocols as geodesics (shortest paths) between endpoints in control-parameter space that minimize resistance to driving. 

The generalized friction captures local features of the system's free energy and dynamic relaxation throughout collective-variable space; these features are also relevant to characterizing transient dynamics during a spontaneous transition path~\cite{Berkowitz1983,Maragliano2006,Zhao2010,Johnson2012}. Intuitively, if the system must overcome a free-energy barrier during the reaction, the spontaneous transitions are likely to pass through a relatively low-free-energy region of collective-variable space to reduce heat absorption during the transient dynamics. It seems similarly intuitive that a minimum-work protocol would drive the system through the same low-free-energy region to reduce the work done that increases the system's energy. This leads us to hypothesize that protocols designed to minimize frictional resistance may also drive the system along the same configuration-space pathways favored by spontaneous transitions. This hypothesis is supported by Ref.~\cite{Rotskoff2017}, where a minimum-work protocol designed to invert the magnetization of a large 2D Ising model showed strong correspondence with the spontaneous transition pathways characterized by a minimum-free-energy path~\cite{Venturoli2009}.

Theoretical descriptions of minimum-work protocols and spontaneous transitions differ in several ways. Minimum-work protocols are performed in a fixed duration with endpoints defined in control-parameter space, while spontaneous transition paths occur in variable duration with fixed endpoints in configuration space. Protocols exchange work with the system throughout the protocol and drive it out of equilibrium, while there is no work performed on the system during a spontaneous transition that occurs when the system is at equilibrium. 

Nevertheless, the processes share some common features that suggest deeper connections. Both processes share the same configuration space, with internal energy coupling the system's many degrees of freedom. With appropriate choice of control parameters, the endpoint distributions of the protocol can approximate the respective unperturbed distributions in metastable basins surrounding transition-path endpoints, providing at minimum a control-parameter space capable of distinguishing metastable conformations. Additionally, reweighting observations of system state throughout the protocol using excess work allows estimation of equilibrium properties of the unperturbed system (such as the potential of mean force)~\cite{Crooks2000,Hummer2001}, which also yields insight into the spontaneous transition mechanisms~\cite{Engel2014}.

The choice of control parameters and manner of driving a system affects the efficiency of minimum-work protocols and estimation of equilibrium properties~\cite{Gore2003,Shenfeld2009,Kim2012,Blaber2020,Blaber2022}, and similarly the choice of collective variables affects the information that can be gained about a spontaneous transition~\cite{Peters2016,Bolhuis2015,Lin2018}. Physically intuitive optimization criteria for characterizing spontaneous transitions are still in development~\cite{Peters2017q}, and we are interested in whether minimizing work in an appropriately chosen control-parameter space can provide a thermodynamic criterion for optimizing paths describing a spontaneous transition. Ultimately, we aim to find some correspondence between minimum-work protocols and spontaneous transitions: how could minimum-work protocols be used to learn about the spontaneous transition mechanism, and how could the spontaneous transition mechanism be used to design efficient protocols? 

We examine the spin-inversion mechanism in a 3$\times$3 Ising model (Fig.~\ref{fig:Ising_Model_2},~\cite{Louwerse2022b}), comparing the transition mechanism and its energetic cost during a spontaneous transition and a protocol designed to minimize work. We find that the two processes show similar orders of spin flips and corresponding internal energy flows to the system, suggesting that designed protocols using the generalized friction metric capitalize on some important system features that correspond to the spontaneous transition mechanism.

\section{Model system and theoretical background}

\begin{figure}
    \centering
    \includegraphics[width=0.9\linewidth]{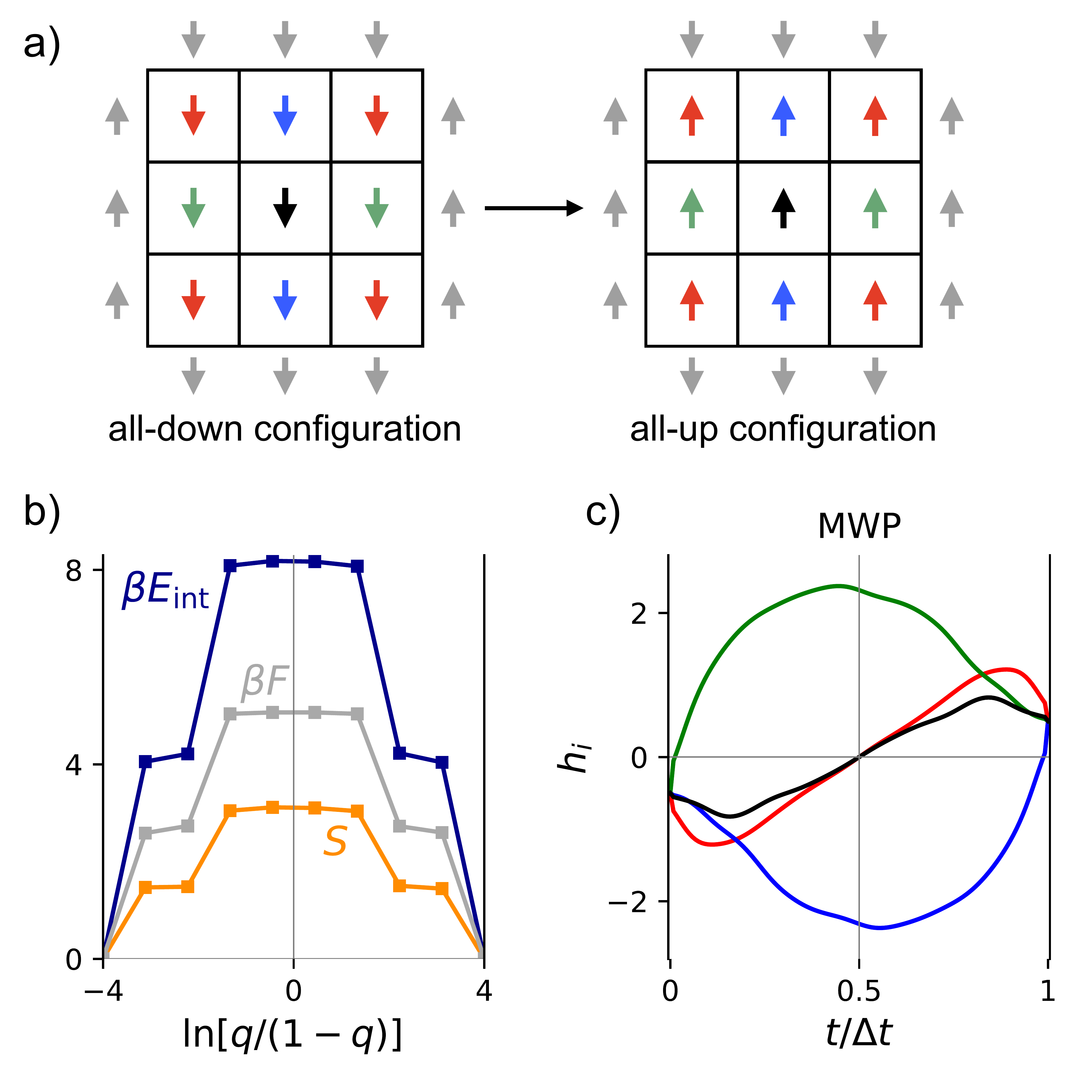}
    \caption[Schematic of spin inversion in 3$\times$3 Ising model.]{a) Schematic of 3$\times$3 Ising model with 9 fluctuating spins (colors) and 12 fixed boundary spins (gray), shown in metastable all-down and all-up configurations. Spins are colored according to their symmetry type. b) Change in mean internal energy (blue), entropy (orange), and free energy (grey) of the system as a function of reaction coordinate $\ln[q/(1-q)]$ during the transition-path ensemble. c) Designed protocol~\cite{Louwerse2022b} for driving spin inversion using four magnetic fields corresponding to colors in a).}
    \label{fig:Ising_Model_2}
\end{figure}

We study a 3$\times$3 Ising model with fixed anti-symmetric boundary conditions~\cite{Louwerse2022b}, illustrated in Fig.~\ref{fig:Ising_Model_2}a. The spins are ferromagnetically coupled, with spin configuration $\st$ having internal energy
\begin{equation} \label{eq:system_energy}
    E_{\rm int}(\st) \equiv -J \sum_{\{ i,j \}} \sigma_i \sigma_j \>,
\end{equation}
where $J=1\,\kT$ is the coupling coefficient for Boltzmann constant $k_{\rm B}$ and temperature $T$, $\sigma_i \in \{ -1, 1 \}$ is the orientation of spin $i$, and $\sum_{\{ i,j \}}$ denotes a sum over nearest-neighbor spin pairs. The probability that the system is in state $\st$ evolves 
according to
the Master equation
\begin{equation} \label{eq:Master_equation}
    \md_t p(\st) = \sum_{\st'} T_{\st \st'} p(\st') \ .
\end{equation}
The $\st' \to \st$ transition rates obey single-spin-flip Glauber dynamics~\cite{Glauber1963}, 
\begin{equation}
T_{\st \st'} = \frac{1}{9} \frac{1}{1+e^{\beta [ E_{\rm int}(\st) -E_{\rm int}(\st')]}} \ ,
\end{equation}
with $T_{\st \st} = - \sum_{\st'} T_{\st' \st}$. The prefactor is in units of inverse attempted spin flips and the Glauber acceptance probability enforces detailed balance~\cite{Zwanzig2001}.

\subsection{Transition-path ensemble for the Ising model}

The Ising system has two energetically stable configurations, with spins either all down or all up, which are chosen as the endpoints of the reaction (Fig.~\ref{fig:Ising_Model_2}a). The spontaneous transition is described by the transition-path ensemble~\cite{E2006,Metzner2009}, the set of all trajectories that transit from all-down to all-up without visiting either state in between. The transition-path ensemble is characterized by the committor $q(\st)$, the probability that a trajectory initiated from microstate $\st$ reaches the all-up configuration before returning to the all-down configuration. It can be calculated in discrete systems by solving the recursion relation~\cite{Metzner2009}
\begin{equation}
    0 = \sum_{\st'} T_{\st' \st} q(\st') \ ,
\end{equation}
with boundary conditions $q(\st_{\rm d})=0$ for $\st_{\rm d}$ the all-down configuration and $q(\st_{\rm u})=1$ for $\st_{\rm u}$ the all-up configuration. The committor projects the state space onto the unit interval, $q(\st) \in [0,1]$, and serves as a reaction coordinate describing the transition, also allowing calculation of various reaction properties~\cite{Metzner2009,Vanden-Eijnden2014}. 

For the Ising system, the transition-state ensemble (set of configurations with $q(\st) \approx 0.5$) poses an internal-energy barrier (compared to the configuration with $q(\st)=0$) of $8\, \kT$ and has entropy $\approx 3 \, k_{\rm B}$, combining to yield a free-energy barrier of $\approx 5\, \kT$ (Fig.~\ref{fig:Ising_Model_2}b). Due to the system symmetry, there is no energy difference between reaction endpoints. The mean first-passage time for the reaction is $\approx 1890$ attempted spin flips, and the mean transition-path duration is $\approx 89$ attempted spin flips~\cite{Vanden-Eijnden2014}.

The system dynamics in the transition-path ensemble satisfy modified transition rates~\cite{Vanden-Eijnden2014,Louwerse2022a}. A trajectory is initialized in the all-down configuration $\st_{\rm d}$, then transitions to $\st$ with probability
\begin{equation} \label{eq:TPE_first_step}
    p^{\rm initial}_{\rm R}(\st) = \frac{T_{\st \st_{\rm d}} q(\st)} {\sum_{\st''} T_{\st'' \st_{\rm d}} q(\st'')} \>,
\end{equation}
where the denominator normalizes over all possible transitions out of the all-down configuration, and subscript $\rm{R}$ indicates the reactive transition-path ensemble. Subsequent transitions in the forward transition-path ensemble obey
\begin{equation} 
	\label{eq:Ising_TPE_rates}
	T^{\rm R}_{\st \st'} = T_{\st \st'} \frac{q(\st)}{q(\st')} \>,
\end{equation}
until the system reaches the all-up configuration. These modified transition rates are used to directly generate an ensemble of transition paths. The transition-path ensemble has a steady-state probability distribution,
\begin{equation} \label{eq:TPE_cond_distr}
    p_{\rm R}(\st) = \frac{\pi(\st)q(\st)[1-q(\st)]}{p_{\rm R}} \>,
\end{equation}
where 
\begin{equation}
    \pi(\st)=e^{\beta[F-E_{\rm int}(\st)]}
\end{equation} 
is the equilibrium probability of $\st$ with free energy $F=-\kT \ln \sum_{\st} e^{-\beta E_{\rm int}(\st)}$, and 
\begin{equation}
    p_{\rm R} = \sum_{\st} \pi(\st)q(\st)[1-q(\st)]
\end{equation} 
is the probability a system at equilibrium is currently on a reactive trajectory, which normalizes the state distribution (Eq.~\ref{eq:TPE_cond_distr}) during the transition-path ensemble.

\subsection{Minimum-work protocol for Ising model}

In addition to the spontaneous transition, we consider driving the spin inversion using a set of time-dependent applied magnetic fields $\bm{h}(t)$ 
imposing external energy
\begin{equation} \label{eq:control_energy}
    E_{\rm ext}(\st,\bm{h}) = -\bm{h}^{\rm T} \cdot \bm{X}(\st) \>,
\end{equation}
{\ML the product of each magnetic field $h_i$ with}\stkout{where $X_i(\st)$ is} the total magnetization {\ML  $X_i(\st)$} of spins {\ML it controls.} \stkout{controlled by field $h_i$.} Thus the total energy is
\begin{equation}
E_{\rm tot}(\bm{\sigma},\bm{h})=E_{\rm int}(\bm{\sigma})+E_{\rm ext}(\bm{\sigma},\bm{h}) \ ,
\end{equation}
{\ML the sum of the internal energy $E_{\rm int}(\bm{\sigma})$ arising from the coupling between spins [Eq.~\eqref{eq:system_energy}] 
and external energy $E_{\rm ext}(\bm{\sigma},\bm{h})$ arising from the system-controller coupling [Eq.~\eqref{eq:control_energy}]. During a spontaneous transition, the system is in equilibrium and external energy is zero ($\bm{h}=\bm{0}$), while during a control protocol, work is done on the system by changing the external energy through changes in fields $\bm{h}$.}

The spin magnetizations serve as both collective variables used to describe the spontaneous transition and as conjugate forces $X_i = -\partial E_{\rm tot} / \partial h_i$ to the control parameters. The system magnetization is inverted by changing the magnetic fields from $h_i(t=0)=-0.5\,\kT$ for all $i$ (favoring the all-down configuration) to $h_i(t=\Delta t) = 0.5\,\kT$ (favoring the all-up configuration). 

For long-duration protocols, the excess power to the system at time $t$ is approximated using linear-response theory~\cite{Sivak2012} as
\begin{equation} \label{eq:excess_power}
	\langle P^{\rm ex} (t) \rangle_{\Lambda} \approx \sum_{i j} \dot{h}_i(t) \zeta_{i j}(\bm{h}(t)) \dot{h}_j(t) \>,
\end{equation}
where subscript $\Lambda$ indicates an average over the control protocol. Here, $\bm{\zeta}(\bm{h})$ is the generalized friction metric at $\bm{h}$, given by the integral of the temporal correlation function between conjugate forces, 
\begin{equation}
	\zeta_{i j}(\bm{h}) \equiv \int_0^{\infty} \md t \langle \delta X_i(0) \delta X_j(t) \rangle_{\bm{h}} \>.
\end{equation}
The generalized friction is a Riemannian metric that provides a measure of distance between equilibrium ensembles in control-parameter space\stkout{.}{\ML, where the mean excess work
\begin{equation} \label{eq:LR_excess_work}
\langle W^{\rm ex} \rangle \approx \int_0^{\Delta t} \md t \, \bm{h}^{\rm T}(t) \cdot \bm{\zeta}(\bm{h}(t)) \cdot \bm{h}(t)
\end{equation}
is related to the ``length'' of the protocol curve in control-parameter space.}
This geometric interpretation has implications for our understanding of minimum-work protocols: minimum-work protocols in the long-duration limit are geodesics (shortest paths) between control-parameter endpoints, which implies that they are independent of protocol duration. The linear-response approximation therefore allows us to find minimum-work protocols for any protocol duration sufficiently long that the system is in the linear-response regime. In previous work~\cite{Louwerse2022b}, we \stkout{calculated}{\ML used the string method to numerically solve the Euler-Lagrange equation for Eq.~\eqref{eq:LR_excess_work}, which identifies the} \stkout{a} minimum-work protocol (Fig.~\ref{fig:Ising_Model_2}c) that drives the system using four magnetic fields corresponding to the four colors in Fig.~\ref{fig:Ising_Model_2}a\stkout{.}{\ML, a relatively low-dimensional control-parameter space that biases all spins while preserving the symmetry of the boundary spins.} Here, we compare the minimum-work protocol to the spontaneous transition mechanism.

\section{Transition mechanism for spin inversion}

\subsection{Reaction endpoints}

\begin{figure}
	\centering
	\includegraphics[width=\linewidth]{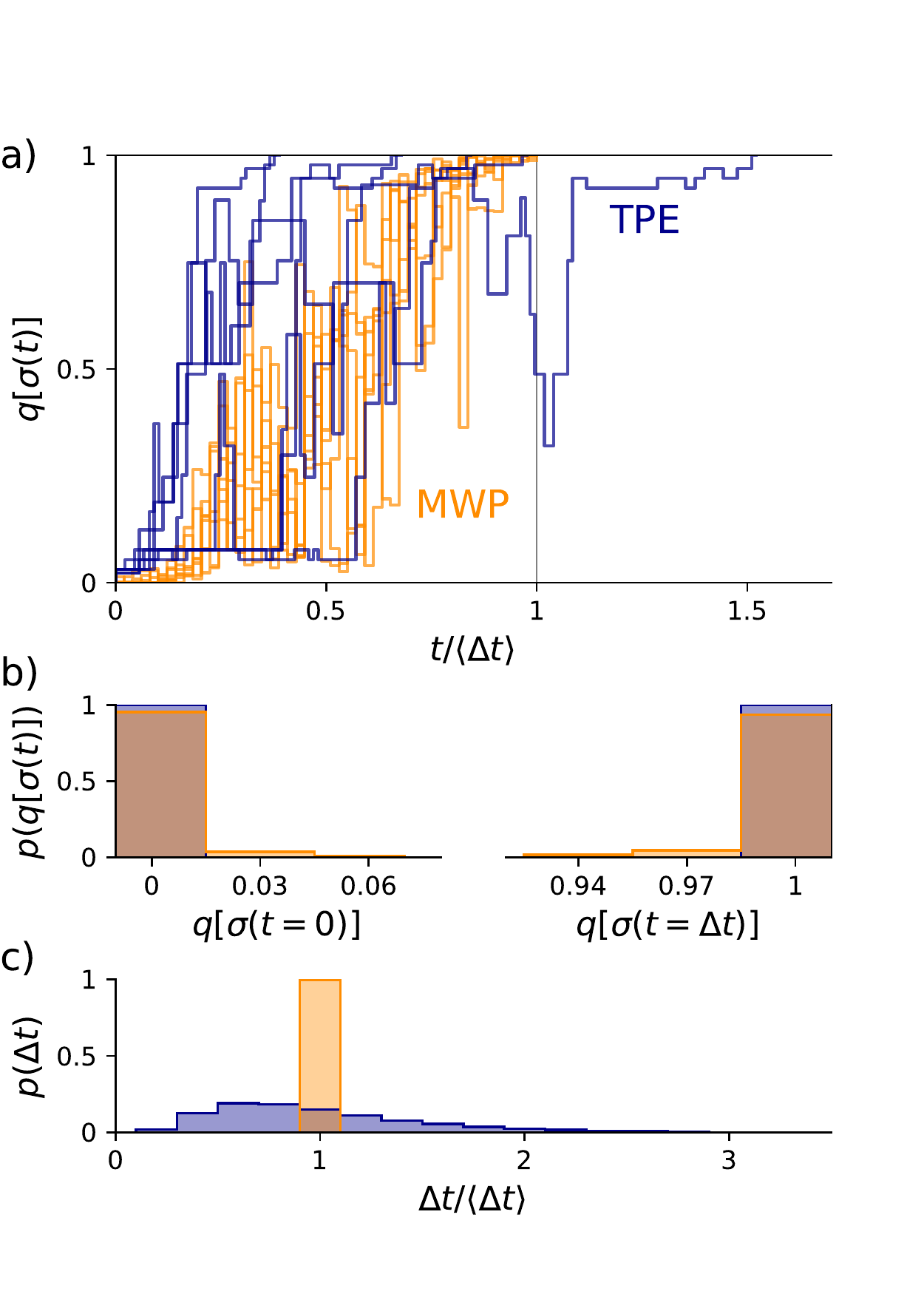}
	\caption[Sample trajectories from MWP and TPE.]{Trajectory ensembles from the minimum-work protocol and transition-path ensemble. a) 7 sample trajectories generated by a minimum-work protocol (orange) and from the transition-path ensemble (dark blue), projected onto the committor $q(\st)$. The trajectory time is scaled by the mean duration $\langle \Delta t \rangle$ of trajectories in the respective ensemble. b) Distribution of start and end configurations for trajectories in each ensemble. c) Distribution of trajectory durations for each ensemble.}
	\label{fig:traj_distr_MWP_TPE}
\end{figure}

The ensemble of system trajectories generated by a minimum-work protocol are equal in duration and differ in their start and end configurations, while trajectories making up the transition-path ensemble have the same endpoints (the all-down and all-up configurations) but differ in duration, as illustrated in Fig.~\ref{fig:traj_distr_MWP_TPE}.

The control parameters at the protocol endpoints are chosen to restrain the system (when at equilibrium) to the all-down and all-up configurations respectively. The system begins the protocol in equilibrium (dominated by the all-down configuration), but does not reach the equilibrium distribution by the end of a finite-duration protocol; for the sufficiently long protocol considered here ($\Delta t = 1000$ attempted spin flips~\cite{Louwerse2022b}), at the protocol's end the system will be close to equilibrium and therefore primarily occupy the all-up configuration. Figure~\ref{fig:traj_distr_MWP_TPE}b shows that the distribution of committor values for configurations at the start and end of the designed protocol of this duration are highly peaked in the $q=0$ and $q=1$ states respectively, indicating that the trajectories generated by a protocol of this duration overwhelmingly transit between the reaction endpoints.

\subsection{Reaction coordinate}

Characterizing the system trajectory between the protocol endpoints requires a parameterization of configurations along the trajectory. The natural one-dimensional parameterization for the transition-path ensemble is the committor~\cite{Peters2016,Bolhuis2015}; in contrast, for minimum-work protocols the scaled protocol duration $t/\Delta t \in [0,1]$ is a natural parameterization since all trajectories are of equal duration and the control parameters have the same value at time $t$ for all repetitions of the protocol, providing a similar force to the system at that time. We use both methods of parameterizing the trajectory ensembles, analyzing trajectories as a function of trajectory time and as a function of the committor.

We divide the range of committor values into discrete bins, grouping together all configurations from each trajectory that fall in the same bin. Since the committor typically increases rapidly around an energy barrier~\cite{Li2015}, bins spaced linearly in the committor coordinate tend to under-emphasize the variation of configurations at the start and end of the transition mechanism. To make this variation more visible, we use a nonlinear transformation of the committor, 
\begin{equation}
    \logit(\st) = \ln \frac{q(\st)}{1-q(\st)} \>.
\end{equation}
Such an invertible transformation of the committor does not affect the information contained in the one-dimensional coordinate, so $\logit(\st)$ remains a sufficient reaction coordinate for characterizing reaction details~\cite{Louwerse2022a}. 

The average of arbitrary observable $A$ at transformed committor value $\logit_0$ is
\begin{equation}
    \langle A \rangle_{\logit_0} = \frac{\langle A \, \bm{1}_{\logit_0}(\logit) \rangle}{\langle \bm{1}_{\logit_0}(\logit) \rangle} \ ,
\end{equation}
for the indicator function
\begin{equation}
\bm{1}_{\logit_0}(\logit) =
    \begin{cases}
    1 \quad & \logit \in [\logit_0 \pm \Delta \logit/2 ) \\
    0 \quad & \rm{otherwise}
    \end{cases}
\end{equation}
that selects states in the bin of width $\Delta \logit$ centered at $\logit_0$. Angle brackets denote an average over all states in the appropriate trajectory ensemble,
\begin{equation}
    \langle A \rangle = \frac{1}{N_{\rm traj}} \sum_{n=1}^{N_{\rm traj}} \frac{1}{ M^{(n)}} \sum_{m=0}^{M^{(n)}} A^{(n)}(t_m) \>,
\end{equation}
for $N_{\rm traj}$ trajectories in the sample, $M^{(n)}$ attempted spin flips in the $n$th trajectory, and $A=A^{(n)}(t_m)$ at time $t_m$ during the $n$th trajectory. 

We also use the scaled trajectory time $\tau_0 \in [0,1]$ to parameterize the ensembles, the proportion $t/\Delta t$ of time elapsed in the current trajectory of duration $\Delta t$. The conditional mean in each bin is first determined for a single trajectory, 
\begin{equation}
    \langle A^{(n)} \rangle_{\tau_0} = \frac{\sum_{m=0}^{M^{(n)}} A^{(n)}(t_m) \bm{1}_{\tau_0}(t_m/\Delta t) }{\sum_{m=0}^{M^{(n)}} \bm{1}_{\tau_0}(t_m/\Delta t)] } \>.
\end{equation}
This mean is then averaged over all trajectories to ensure that longer trajectories with more states in each bin do not dominate the mean~\cite{Cossio2018},
\begin{equation}
    \langle A \rangle_{\tau_0} = \frac{1}{N_{\rm traj}} \sum_{n=1}^{N_{\rm traj}} \langle A^{(n)} \rangle_{\tau_0} \>.
\end{equation}
This averaging ensures that each trajectory has equal weight in the ensemble average, despite the variety of trajectory durations in the transition-path ensemble (Fig.~\ref{fig:traj_distr_MWP_TPE}c).

\subsection{Symmetry-breaking of trajectories}

\begin{figure} 
    \centering
    \includegraphics[width=\linewidth]{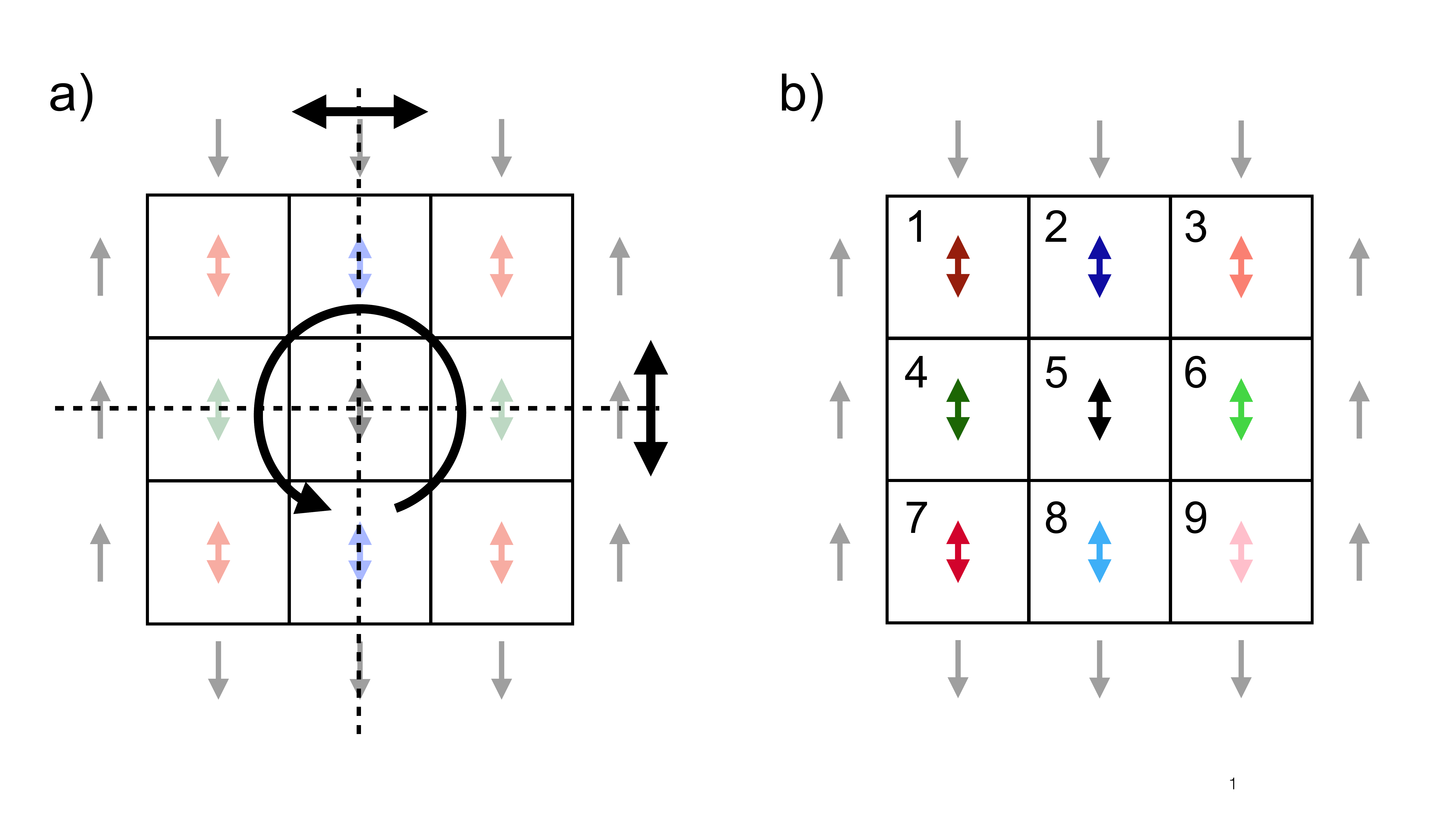}
    \caption{a) Symmetry operations for 3$\times$3 Ising model: a horizontal reflection, a vertical reflection, and a $180^{\rm o}$ rotation. b) Color-coding for 9 different spin types.} 
    \label{fig:symmetry_transformations}
\end{figure}

The 4D minimum-work protocol preserves the symmetry imposed by the boundary conditions, but there is no guarantee that the transition-path ensemble preserves this symmetry. We therefore perform symmetry operations (Figure~\ref{fig:symmetry_transformations}) on each trajectory to resolve differences in the flip timing of spins of the same color that would otherwise be obscured by the system's underlying symmetry.

For each trajectory, spins are initially ordered according to the proportion of time spent in the up orientation, where $\ell_i$ is an integer indicating the order in which spin $i$ flips during the trajectory. A symmetry operation is chosen that minimizes
\begin{equation}
    \sum_{i \in [1,2,4]} \ell_i-\sum_{j \in [6,8,9]} \ell_j
\end{equation}
to generally place spins that flip earlier in the trajectory in the upper left corner and spins that flip later in the lower right. (Numbers in each summation indicate the corresponding spins in Fig.~\ref{fig:symmetry_transformations}b.) The chosen symmetry operation is applied to the entire trajectory, then mean properties of each spin are averaged over all trajectories. 

\subsection{Transition mechanism}

\begin{figure}
    \centering 
    \includegraphics[width=\linewidth]{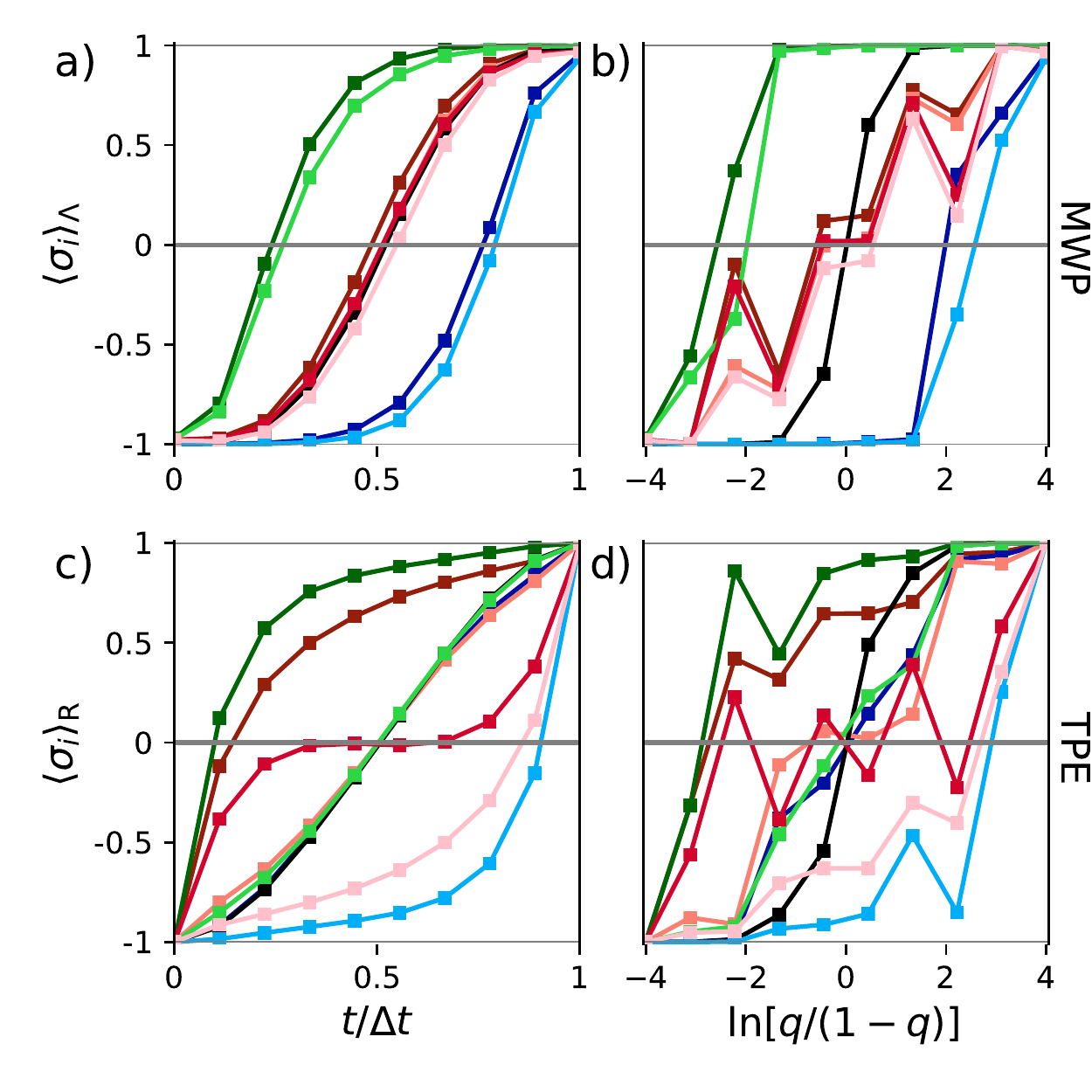}
    \caption[Mean magnetization of spin sets during MWP and TPE.]{Mean magnetization of each spin (colors in Fig.~\ref{fig:symmetry_transformations}b) at (a,c) scaled trajectory time $t/ \Delta t$ and (b,d) transformed committor $\ln [q/(1-q)]$ during the (a,b) minimum-work protocol and (c,d) transition-path ensemble.}
    \label{fig:per_spin_mechanism}
\end{figure}

Figure~\ref{fig:per_spin_mechanism} shows the mean state of each spin set during the 4D fully optimized protocol and during the transition path as parameterized by the scaled transition-path time and transformed committor. The two methods of parameterizing the reaction show qualitative similarities in the timing of changes to each spin's mean state. The mean system state during the 4D fully optimized protocol and during the transition-path ensemble show similar characteristics. In general, the green spins (which are initially energetically frustrated due to the adjacent boundary spins of opposite sign) flip relatively early in both processes (i.e., cross zero average magnetization for $t/\Delta t<0.5$ and $\ln[q/(1-q)]<0$), and the blue spins (which end in an energetically frustrated orientation) flip relatively late. The red and black spins flip throughout the middle of the protocol. 

However, the finer details of the transition mechanism differ between the transition-path ensemble and minimum-work protocol. In the minimum-work protocol, both green spins flip early, then all red and black spins flip, then finally both blue spins. Spins in the same spin set flip at approximately the same time, reflecting the symmetry of the driving protocol. In contrast, the transition-path ensemble shows symmetry breaking for spins in the same spin set: a green and red spin flip first; followed by a second red spin; then the second green, first blue, and black spins; then the third red spin; and finally the fourth red and second blue spins. 

The differences in the detailed mechanisms are likely due to the constraint on driving in the 4D control parameter space where, e.g., all four red fields are changed in the same way throughout the protocol. With this constraint, it is not possible for the protocol to drive one spin to the up orientation independently of other spins in the same spin set. The resulting protocol in the 4D space is therefore symmetric in the identity of each of the spins, whereas this symmetry is broken in the transition-path ensemble. However, the general feature of green spin(s) flipping early and blue spin(s) flipping late is preserved in both processes, reflecting the bias imposed by the fixed boundary spins that affects both the geometry of control-parameter space and the spontaneous transition mechanism.

\begin{figure} 
    \centering
    \includegraphics[width=\linewidth]{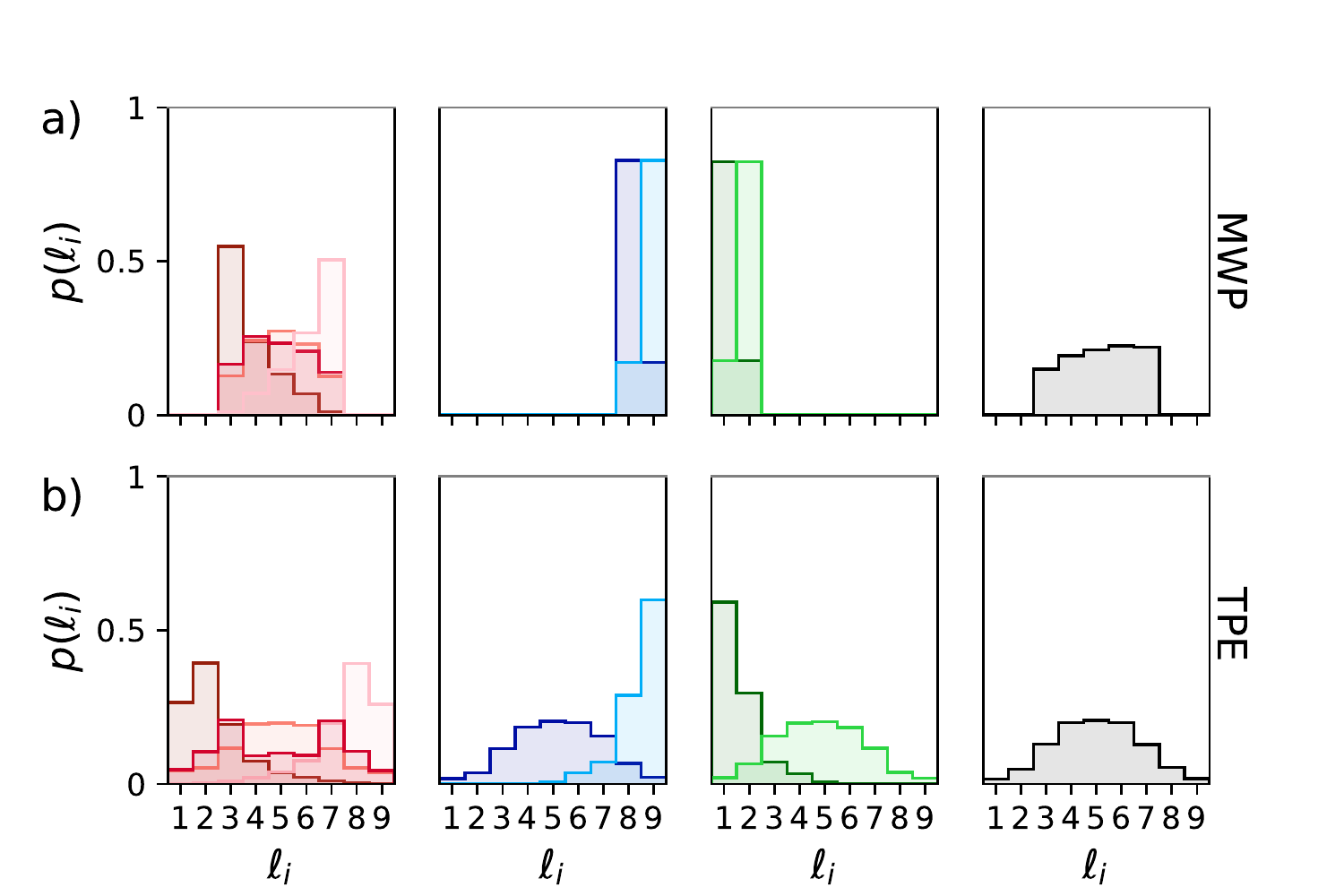}
    \caption[Distribution of spin-flip order during the MWP and TPE.]{Distribution of spin-flip order $\ell_i$ for each spin type during the a) minimum-work protocol and b) transition-path ensemble. Each histogram shows the distribution of the spin-flip order for a given spin, with spins in the same set separated based on their flipping order in the trajectory. Colors of each histogram correspond to spins in Fig.~\ref{fig:symmetry_transformations}b.}
    \label{fig:order_distributions}
\end{figure}

Figure~\ref{fig:order_distributions} shows the overall order of spin flips during the trajectory by plotting the distribution of the spin-flip order $\ell_i$ of each spin type. The minimum-work protocol preserves the order across all trajectories, with both green (blue) spins flipping first (last), and red and black spins all flipping in between. (The symmetry operations that attempt to place the first spin flips in the upper-left corner and last spin flips in the bottom right does not accomplish this perfectly for the minimum-work protocol: in approximately $20 \%$ of trajectories, the light green (light blue) spin flips before the dark green (dark blue).)

The transition-path ensemble spin-order distributions show higher diversity. The upper-left red and green spins predominantly flip first, and the lower-right blue and red spins predominantly flip last, but the order of other spins is more variable. This indicates that there are many paths taken (with significant probability) with differing spin-flip orders. It is possible that the transition-path ensemble mechanism in this system cannot be characterized by a single ordering of spin flips, and instead some other physically intuitive collective variables may better summarize the variety of transition paths.

\section{Energy flows}

\subsection{Theoretical description}

Control protocols produce changes in the system configuration by doing work. Transition paths overcome an energy barrier by waiting for a sequence of fluctuations from the environment that provide the necessary energy as heat to the system. It seems intuitive that if the system must overcome the same internal energy barrier while changing between collective-variable endpoints, then a ``good'' control protocol would provide this energy to the system as work rather than waiting for the appropriate (rare) thermal fluctuations from the environment, essentially flattening the total energy landscape along the path the system takes through configuration space. Here, we analyze the energy flows into the system during both processes. 
{\ML Note that the equilibrium ensemble has no net heat flow; however, the ensemble of forward transition paths breaks time-reversal symmetry (the time-reverse of each forward transition path is a reverse transition path that is not included in the calculation) and so permits net heat flow.}
Analyzing energy flows in the transition-path ensemble has been discussed previously~\cite{Li2016,Wu2022} where energy flows are shown to be helpful for determining collective variables that may be relevant to the reaction and analyzing their role in the mechanism.

The heat flow to the system is the total energy change when the system changes state. For a transition path, the heat flow is the change in internal energy
\begin{subequations} \label{eq:heat_TPE}
\begin{align}
    Q_{\rm R} &= Q_{\rm int} \\
    &= \Delta E_{\rm int} \>,
    \end{align}
\end{subequations}
while the heat flow for a control protocol changes both the internal {\ML [Eq.~\eqref{eq:system_energy}]} and external energy {\ML [Eq.~\eqref{eq:control_energy}]}
\begin{subequations} \label{eq:heat_protocol}
\begin{align}
    Q_{\Lambda} &= Q_{\rm int} + Q_{\rm ext} \\ 
    &= \Delta E_{\rm int} + \Delta E_{\rm ext} \>.
    \end{align}
\end{subequations}
Control protocols and transition paths share the internal-energy landscape $E_{\rm int}$. Therefore we analyze changes in internal energy during both processes to better understand how the system overcomes the (fixed) internal-energy barrier. Additionally, since we hypothesize that the control parameters provide the necessary energetic bias to push the system into high-internal-energy configurations, we calculate the change in external energy for each spin and compare to the changes in internal energy during the minimum-work protocol.

Since the single-spin-flip dynamics are multipartite (where only one spin flips in a given time step while all others remain stationary)~\cite{Hartich2014,Horowitz2015}, the total heat flow can be split into contributions from each spin to better understand the energetic costs associated with each spin's dynamics during the mechanism. The mean heat flow to spin $\sigma_i$ during a trajectory is
\begin{subequations}
\begin{align}
    \langle Q^i \rangle_{\logit_0} =& \\
    \frac{1}{N_{\rm traj}} &\sum_{n=1}^{N_{\rm traj}} \sum_{m=0}^{M^{(n)}-1} Q^i_{\st^{(n)}_{m+1}, \st^{(n)}_m} \bm{1}_{\logit_0}(\tfrac{1}{2} [\logit( t_{m+1}) + \logit(t_m)] ) \nonumber \\
    \langle Q^i \rangle_{\tau_0} =& \frac{1}{N_{\rm traj}} \sum_{n=1}^{N_{\rm traj}} \sum_{m=0}^{M^{(n)}-1} Q^i_{\st^{(n)}_{m+1}, \st^{(n)}_m} \bm{1}_{\tau_0} \left( \frac{\tfrac{1}{2}[t_{m+1}+t_m]}{\Delta t} \right) 
\end{align}
\end{subequations}
where $Q^i_{\st^{(n)}_{m+1}, \st^{(n)}_m}$ is the heat flow (Eq.~\eqref{eq:heat_TPE} for the transition-path ensemble and Eq.~\eqref{eq:heat_protocol} for the minimum-work protocol) due to spin $i$ flipping in step $t_m \to t_{m+1}$ during the $n$th trajectory (equaling zero if spin $i$ does not flip).
If heat flow is positive into spin $i$, i.e. $\langle \Delta E_{\rm int}^{i} \rangle > 0$, this suggests that spin $i$ flipping primarily ``activates'' the system towards the energy barrier, while $\langle \Delta E_{\rm int}^{i} \rangle < 0$ implies that flipping of spin $i$ primarily relaxes the system to the product~\cite{Li2016}.

\subsection{Energy flows to spins}
\begin{figure*}
    \centering
    \includegraphics[width=\linewidth]{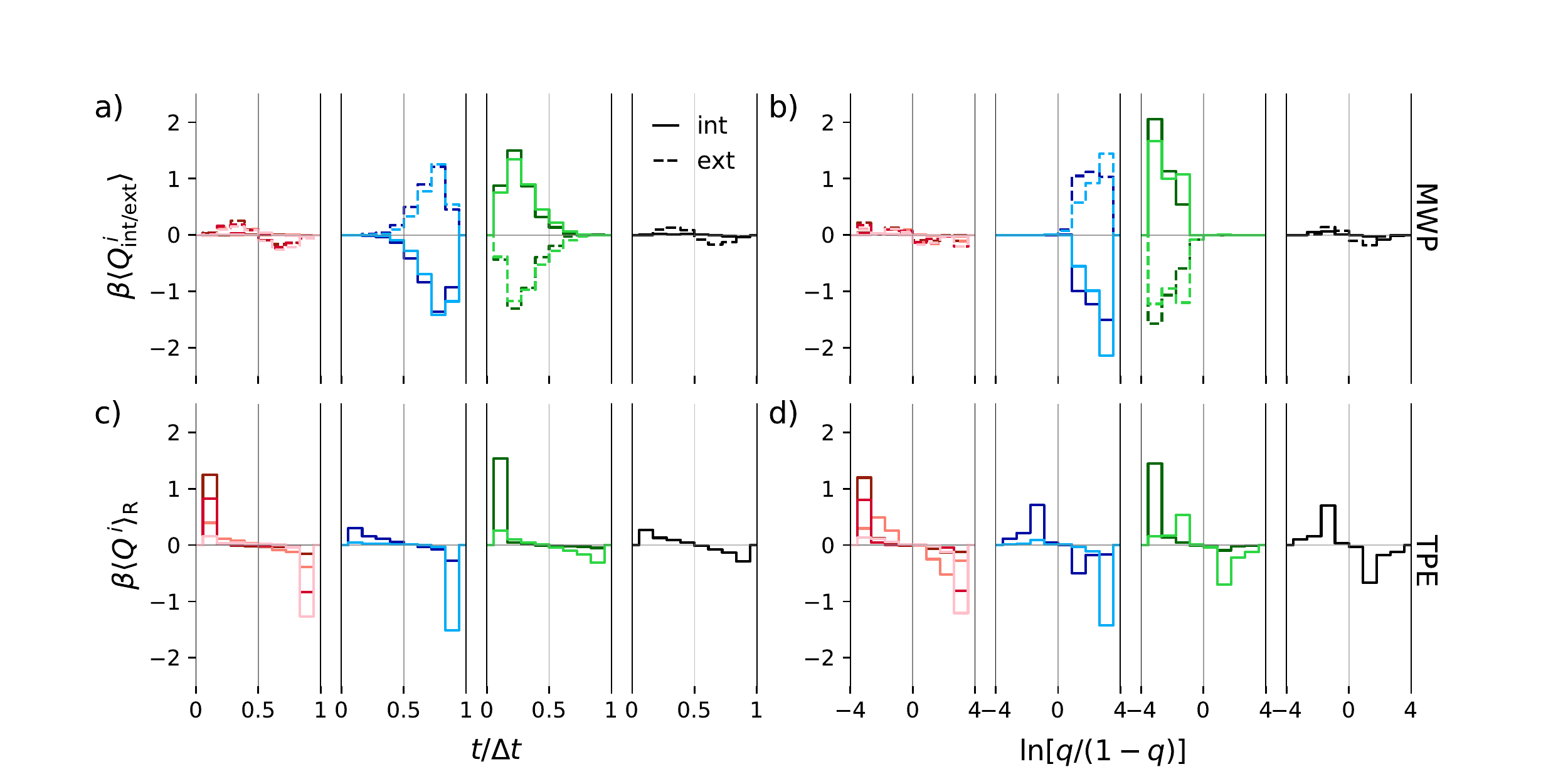} 
    \caption[Heat flow to each spin during the MWP and TPE.]{Heat flow to each spin during the minimum-work protocol and the transition-path ensemble. (a,b) External heat flow $\beta \langle Q^i_{\rm ext} \rangle$ (dashed) and internal heat flow $\beta \langle Q^i_{\rm int} \rangle$ (solid) due to magnetization change in $\sigma_i$ during the minimum-work protocol as a function of a) scaled trajectory time $t/\Delta t$ and b) transformed committor $\ln[q/(1-q)]$. (c,d) Internal heat flow $\beta \langle Q^i \rangle_{\rm R}$ due to magnetization change in $\sigma_i$ during the transition-path ensemble as a function of c) scaled trajectory time $t/\Delta t$ and d) transformed committor $\ln[q/(1-q)]$.}
    \label{fig:per_spin_energy}
\end{figure*}

Figure~\ref{fig:per_spin_energy} shows the mean heat flow during the minimum-work protocol and during a transition path. When the system moves during the minimum-work protocol, some of the heat flow changes internal energy and some changes external energy (Fig.~\ref{fig:per_spin_energy}a,b). The change in internal energy for the green spins is positive during the first half of the protocol, for the blue spins is negative during the second half of the protocol, and for red and black is zero throughout the protocol. The changes in internal energy for the blue and green spins are compensated by changes in the external energy: the external energy decreases on average when green spins flip (increasing the internal energy) and increases on average when blue spins flip (decreasing internal energy). This demonstrates how external energy is provided to the system, allowing it to access high-internal-energy configurations during the protocol. When control parameters do work they change only the external energy; therefore the energetic bias driving the system to access high-internal-energy configurations is provided by control parameters. Additionally, spins of the same type have nearly identical energy flows during the protocol, again reflecting the symmetry of the control parameters.

Figures~\ref{fig:per_spin_energy}c,d show the heat flow (change in internal energy) during the transition-path ensemble for both scaled transition-path time and transformed committor. The symmetry-breaking observed in the reaction mechanism (Fig.~\ref{fig:per_spin_mechanism}) is also reflected in the changes in internal energy during the transition path. The first green and first red spins take in energy on average early in the transition path, bringing the system out of the all-down configuration into higher-internal-energy configurations. Similarly, at the end of the transition path, the second blue and last red spin release energy as the system reaches the low-internal-energy all-up configuration. Throughout the middle of the protocol, the other spins on average take in or release heat. In general, if a spin flips in the first half of the trajectory, the heat flow is positive; if it flips in the second half of the trajectory, the heat flow is negative.

The heat flow during the transition-path ensemble also demonstrates differences in how the scaled transition-path time and transformed committor provide information about the mechanism. While the mean state in these coordinates looks qualitatively similar (Fig.~\ref{fig:per_spin_mechanism}), it is harder to visualize the heat flows using the scaled transition-path time than the transformed committor. The heat flow into the system during the first step out of the all-down configuration and the heat flow out of the system during the final step into the all-up configuration inevitably occur in the first and last bins of the scaled transition-path time, leading to heat flow being highly peaked at these times. Throughout the middle of the transition path, heat flow is less than at the start and end for two reasons: first, some of the internal-energy changes for transitions throughout the trajectory are zero and thus do not contribute to the heat flow; second, linearly rescaling the transition-path time can result in a specific spin flip occurring at a wide range of scaled transition-path times irrespective of its flip order in the trajectory, which spreads out the heat flow associated with any given spin flip over multiple scaled-trajectory-time bins. In contrast, the transformed committor groups together configurations with similar reaction progress independent of the time that the system takes to reach the state in a given trajectory, and therefore heat flows to the different spins happen at distinct values of the transformed committor. For these reasons, parameterizing system properties in the transition-path ensemble using the transformed committor gives more insight than the scaled trajectory time: the committor is a natural measure of the system's progress between reaction endpoints.

\section{Discussion}

We undertook the first systematic comparison of minimum-work protocols, determined by the generalized friction metric, and spontaneous transition paths in collective-variable space. We found intuitive ways to compare the two conceptually different processes, that reveal qualitative similarities. We have investigated the transition mechanism for spin inversion in a 3$\times$3 Ising model during the fully optimized 4D protocol and during a spontaneous transition path. During the minimum-work protocol, work is done on the system, which provides an energetic bias that drives the system over the energy barrier. In the transition-path ensemble, the system must wait for appropriate fluctuations of heat from the environment to overcome the same barrier.

The minimum-work protocol shows a clear transition mechanism, with both green spins (initially frustrated) taking in energy as they flip first, red and black spins flipping throughout the middle of the protocol with minimal energetic cost, and finally the blue spins (frustrated in the final configuration) flipping and releasing energy. This ordering preserves the symmetry of different spin types and is conserved across all trajectories (Fig.~\ref{fig:order_distributions}). The transition-path ensemble shows a wider diversity of transition paths that break the underlying symmetry of spin types. The first two spin flips are usually one green and one red spin, taking in energy. The next five steps have many orders of spin flips and the final two steps involve flipping a red and blue spin, releasing heat.

The internal energy barrier of $8\, \kT$ is consistent between the two processes, but the overall manner in which the system overcomes the barrier differs. A significant factor affecting the comparison is the constrained symmetry of the 4D protocols, which prevents the controller from pushing on one spin differently than its symmetric counterpart(s). The transition-path ensemble shows symmetry-breaking between spins of the same type, in both the spin-flip order and the energetic cost. To recapitulate this symmetry-breaking during control requires a higher-dimensional protocol. The minimum-work protocol in 9D space is too computationally expensive to compute using the same methods used to generate the 4D fully optimized protocol. Minimum-work protocols using the linear-response approximation in higher-dimensional spaces have been computed (e.g., a 100D protocol in Ref.~\cite{Rotskoff2017}), but under the assumption that relaxation time is constant throughout control-parameter space so that the friction metric can be approximated as proportional to the force covariance matrix, significantly reducing computational cost. The assumption of constant relaxation time throughout control-parameter space does not hold for this system.

{\ML It is also interesting to compare our results in the 3$\times$3 Ising model with previous results in an analogous but larger system~\cite{Rotskoff2017,Venturoli2009}. The minimum-work protocols for both systems are qualitatively similar, with fields near initially anti-aligned edges (for green spins in our model) flipping early in the protocol and fields near initially aligned edges (for blue spins) flipping later in the protocol (the constraint on our control-parameter space is well-suited to reproducing the same mechanism in the smaller model). The spontaneous transitions in our small model are qualitatively similar to the minimum-work protocol despite variation in spins of the same type; in the larger model the correspondence is even stronger, with both the minimum-work protocol and spontaneous transition path flipping all symmetrically situated spins at the same time.}

The choice of control parameters and their ability to effect change in collective variables that are relevant to describing the reaction mechanism is an important factor in the comparison. The 4D protocol studied here cannot reproduce the dynamics in the transition-path ensemble because some of the relevant information is coarse-grained. The Ising model has $2^9=512$ configurations, but only $168$ are unique under the symmetry operations. Each of these has a unique committor value, indicating that the specific geometry of each state is relevant to parameterizing the committor. On the other hand, the 4D magnetization vector (the conjugate force for the control parameters) has only $5 \times 3 \times 3 \times 2=90$ unique values, representing the possible respective magnetizations of the collections of red, green, blue, and black spins. Thus there is loss of information about the transition mechanism in the 4D collective-variable space. {\ML This information loss is analogous to the construction of, e.g., a free-energy surface in a low-dimensional collective-variable space; a poor choice of collective variables may inaccurately reproduce free-energy barriers and transition pathways, yet some relevant information about the transition mechanism can still be gleaned from such studies.} Choosing control parameters that drive all collective variables that are relevant to the reaction would allow a closer comparison with the transition-path ensemble. 

The minimum-work protocol seems to drive the system through a specific set of configurations, providing a clear single transition mechanism and accompanying energy flow. In contrast, for this system the transition-path ensemble has a variety of transition mechanisms which makes it difficult to identify any one dominant path. {\ML In other model systems where}\stkout{If} the transition-path ensemble {\ML is}\stkout{were} such that reactive trajectories tend\stkout{ed} to follow the same path (i.e., lie in a ``transition tube'') and control parameters are chosen to push on the relevant collective variables to describe the transition {\ML (i.e., calculate the committor)}, a stronger correspondence may be observed. Similar comparison of minimum-work protocols in other model systems {\ML and with different functional forms of the external energy} would be valuable for further elucidating the connections between efficient control and spontaneous transition paths.

\vspace{2ex}
\acknowledgments
We thank Steven Blaber (SFU Physics) for stimulating discussion and helpful comments on the manuscript. This work was supported by Natural Sciences and Engineering Research Council of Canada (NSERC) Canada Graduate Scholarships Masters and Doctoral (MDL), an NSERC Discovery Grant (DAS), and a Tier-II Canada Research Chair (DAS). Computational support was provided in part by WestGrid (www.westgrid.ca) and the Digital Research Alliance of Canada (www.alliancecan.ca).


\end{document}